\documentclass[12pt]{article}

\newread\testifexists
\def\GetIfExists #1 {\immediate\openin\testifexists=#1
    \ifeof\testifexists\immediate\closein\testifexists\else
    \immediate\closein\testifexists\input #1\fi}

\usepackage{gthstyle}\usepackage{amsfonts}
\usepackage{amssymb}
\immediate\openin\testifexists=e
    \ifeof\testifexists\immediate\closein\testifexists\else
    \immediate\closein\testifexists\input e\fipsf

\def\bbs#1{\setbox0=\hbox{$\tt #1$}  \copy0\kern-\wd0\kern .15em\copy0}

\def\bbf#1{\setbox0=\hbox{$#1$} \kern-.025em\copy0\kern-\wd0
        \kern.05em\copy0\kern-\wd0 \kern-.025em\raise.0433em\box0}

\def\a{\alpha}      \def\b{\beta}         
\def\d{\delta}        \def\e{\varepsilon}
          \def\l{\lambda}     \def\L{\Lambda}
\def\m{\mu}         \def\f{\phi}    \def\F{\Phi}        
\def\n{\nu}         \def\j{\psi}    
\def\r{\varrho}     \def\s{\sigma}  
\def\t{\tau}        \def\th{\theta}

 \def\LL{{\cal L}} \def\NN{{\cal
N}}
\def\pa{\partial} \def\ra{\rightarrow}

 \def\bel{$\circ$}
\def\dd{{\rm d}}  \def\bra{\langle}   \def\ket{\rangle}

\def\fract#1#2{{\textstyle{#1\over#2}}}
\def\ffract#1#2{\raise .3 em\hbox{$\scriptstyle#1$}\kern-.25em/
                \kern-.2em\lower .2 em \hbox{$\scriptstyle#2$}}

\def\half{\fract12} \def\quart{\fract14} 

\def\part#1#2{{\partial#1\over\partial#2}}

\def\iss{\ =\ }

\newcommand{\tl}[1]{\tilde{#1}}

\newcommand{\be}{\begin{eqnarray}}
\newcommand{\ee}{\end{eqnarray}}
\newcommand{\eqn}[1]{(\ref{#1})}
\newcommand{\nn}{\nonumber\\}

\newcommand{\bi}[1]{\begin{itemize}\item[#1]}
\newcommand{\itm}[1]{\item[#1]}
\newcommand{\ei}{\end{itemize}}

\newcommand{\fn}{\footnote}

\newcommand{\newsec}[1]{\section{#1}\setcounter{equation}{0}}

\def\printversion{\setlength{\textheight}{9in}\setlength{\oddsidemargin}{0in}
    \setlength{\textwidth}{6.3in}\setlength{\topmargin}{-0.1in}}

 \newcommand {\eel}[1]{\label{#1}\end{eqnarray}} \newcommand{\crl}[1]{\label{#1}\\}
   \printversion

\begin{document}
\begin{titlepage}

\title{\normalsize \hfill  ITP-UU-06/25 \\ \hfill SPIN-06/21\\
 \hfill {\tt gr-qc/0606026}\\ \vskip 10mm \Large\bf
THE BLACK HOLE HORIZON AS A DYNAMICAL SYSTEM\thanks{Presented at ``Einstein's Legacy in the New
Millennium", Puri, India, 15-22 Dec. 2005.}}

\author{Gerard 't~Hooft}
\date
 {\normalsize Institute for Theoretical Physics,
Utrecht University \\ and
\medskip \\ Spinoza Institute, Postbox 80.195, 3508 TD
Utrecht, the Netherlands \smallskip \\ e-mail: \tt
g.thooft@phys.uu.nl \\ internet: \tt
http://www.phys.uu.nl/\~{}thooft/}

\maketitle

\begin{quotation} \noindent {\large\bf Abstract } \bigskip

Interactions between outgoing Hawking particles and ingoing matter are determined by gravitational forces and Standard Model interactions. In particular
the gravitational interactions are responsible for the unitarity of the scattering against the horizon, as dictated by the holographic principle, but the
Standard Model interactions also contribute, and understanding their effects is an important first step towards a complete understanding of the horizon's
dynamics. The relation between in- and outgoing states is described in terms of an operator algebra. In this contribution, in which earlier results are
rederived and elaborated upon, we first describe the algebra induced on the horizon by \(U(1)\) vector fields and scalar fields, including the case of an
Englert-Brout-Higgs mechanism, and a more careful consideration of the transverse vector field components.

We demonstrate that, unlike classical black holes, the quantized black hole has on its horizon an imprint of its (recent) past history, i.e., quantum
hair. The relation between in- and outgoing states depends on this imprint. As a first step towards the inclusion of non-Abelian interactions, we then
compute the effects of magnetic monopoles both in the in-states and in the out-states. They completely modify, and indeed simplify, our algebra.

\end{quotation}


\end{titlepage}

\eject
\def\inn{\mathrm{in}}\def\out{\mathrm{out}}\def\intt{\mathrm{int}}
\newsec{Introduction: The black hole scattering matrix}\label{intro.sec}

A growing consensus seems to have been reached concerning the hypothesis that the contribution of black holes to particle scattering phenomena must be
described by a unitary scattering operator\cite{GtH85}\cite{tHSW}\cite{GtH96}\cite{SWH04}, but the agreement is not universal, and indeed, it is not at
all obvious\cite{WUpc}. The author's personal arguments favoring unitarity contain elements that are not at all agreed upon by most of his colleagues;
since these are not essential, we refer to the earlier papers. Whatever one's views are, unitarity, assumed to hold regardless the presence of black
holes, is an essential starting point of this paper, and its sequels.

The unitarity assumption is applied to the region within the Planckian domain surrounding the horizon. The Hawking particles leaving that domain, will
subsequently be affected by the ingoing particles by interactions that we assume to be known. There are two types of interactions that we can take into
account: \bi{\bel} those caused by \emph{perturbative} quantum gravity, and \itm{\bel} those described by some renormalizable quantum field theory, to be
referred to as the ``Standard Model". \ei The gravitational interactions are taken to be perturbative because we are looking at scales where the higher
order corrections may already be assumed to be negligible. They, as well as the Standard Model interactions, can be regarded as ``final state
interactions", taking place well outside the Planckian regime. To handle these final state interactions, the so-called \emph{eikonal
approximation}\cite{AbItz} seems to be appropriate. Outgoing particles scatter against ingoing ones at high center of mass energy and low values for the
exchanged momentum. Possibly, this technique can be used in an entirely general sense, to obtain all effects due to Standard Model interactions, but, in
this paper, we choose a different approach.

The final state interactions are not as innocuous as they might seem. The gravitational ones, in particular, have a divergent Rindler time dependence, so
that their effects are by no means small. Indeed, they modify the local spectrum so much that the unitarity assumption can only be regarded in harmony
with the presence of these effects. Together, the interactions produce a new boundary condition at the horizon, in the form of an operator algebra, which
we wish to study. The picture one obtains appears to be self-consistent, apart from the fact that the transverse components of the gravitational
interactions are difficult to handle. Ignoring the transverse components is tantamount to neglecting the finite size effects in the transverse directions
on the horizon, so that it should not come as a surprise that one then does not recover the desired area law for the statistical entropy.

The transverse components of the gravitational field resemble somewhat the "Standard Model" interactions, and this is why we are taking a closer look at
those first. To make this paper self-sustained, the first part is a repetition of results published earlier\cite{GtH05}; we then continue to handle pure
magnetic charges and draw further conclusions from these new results.

We shall find that operators describing in- and outgoing particles in the vicinity of the horizon are described by an operator algebra. The algebra
represents all interactions between in- and outgoing objects up to some distance \(1/\m\) from the horizon (according to the local metric), where \(\m\)
is the scale parameter at which the Standard Model characteristics have been taken into account.

Consider a small section of the horizon of a black hole, while the hole itself is taken to be large, so that this region of space-time is adequately
described by the Rindler metric. Of course we are in three space dimensions and one time. In locally flat coordinates, the past horizon is described by a
light cone coordinate \(x^+=U_\inn(\tl\s)\) and the future horizon is \(x^-=U_\out(\tl\s)\), where \(\tl\s=(\s_x,\,\s_y)\) are the transverse coordinates
of the horizon. Since (transverse) gravitational interactions are ignored from here on, there will be no back reaction on the metric of the horizon.
Therefore, the horizon's metric can be taken to be a fixed background. In this paper, we limit ourselves to the transverse components of the metric being
\(g_{ij}=\d_{ij}\), so that the \(\tl\s\) are just cartesian coordinates. Generalization to any other choice for the background metric will be
straight-forward.

The main result described in Ref.\cite{GtH96}, is that, in the approximation where transverse gravity is neglected, the ingoing particles can be described
by a momentum distribution \(P_\inn(\tl\s)\) and the outgoing ones by \(P_\out(\tl\s)\), and the following algebra is obtained: \be \big[\,
P_\inn(\tl\s),\,U_\inn(\tl\s')\,\big]&=&\big[\,P_\out(\tl\s),\,U_\out(\tl\s')\,\big]\iss -i\d^2(\tl\s-\tl\s')\crl{pucomm}
\big[\,P_\inn(\tl\s),\,P_\inn(\tl\s')\,\big]&=&\big[\,U_\inn(\tl\s),\,U_\inn(\tl\s')\,\big]\iss 0\ ,\\
\big[\,P_\out(\tl\s),\,P_\out(\tl\s')\,\big]&=&\big[\,U_\out(\tl\s),\,U_\out(\tl\s')\,\big]\iss 0\ ,\\ P_\out(\tl\s)\iss\tl\pa_\s^2\,
U_\inn(\tl\s)\!\!&,&\ P_\inn(\tl\s)\iss -\tl\pa_\s^2\, U_\out(\tl\s)\ .\eel{gravalgebra} This is a gravitational effect; Newton's constant is normalized
to \be 8\pi G=1\ .\eel{newtonnorm} \emph{If} we assume that all states are unambiguously identified by specifying the momentum distribution \(P(\tl\s)\),
we find that we can either choose a basis where \(P_\inn(\tl\s)\) is specified, or a basis where \(P_\out(\tl\s)\) is specified. A unitary transformation
connects these two basis sets, and so we have a unitary \(S\) matrix.\fn{Hawking, in Ref.\cite{Hawking87}, mentions four possible remedies of the black
hole information problem, of which only option \# 3 resembles a bit the view taken here, except that he assumes the black hole information only to
re-emerge just before it collapses completely, a possibility that he rightly rejects. Curiously, the assumption taken here, that information is re-emitted
very shortly after it was absorbed, was not envisaged by him at that time.}

We are interested also in the transverse contribution of gravity, but before handling that, we should obtain experience in deducing such algebras from
interacting field theories; this is why, in this paper, we now concentrate on the Standard Model interactions. One of these was already handled in
Ref.\cite{GtH96}: the Maxwell field interaction. It provides for additional operators, obeying their own algebra. Continuing this way, we get a more
precise operator algebra, which should pave the way to complete control of the horizon.

According to the holographic principle, the variables on the horizon appear to span a Hilbert space associated to a two-dimensional surface, whereas all
in-going and out-going particles in the vicinity of the black hole should be included in these degrees of freedom\cite{GtH93}. This, we actually relax
somewhat. At a given renormalization scale \(\m\) of the Standard Model, we ignore infrared divergences at much lower energy scales. The infra-red cut-off
then amounts to ignoring degrees of freedom far away from the horizon. This way, one may regard our horizon algebra as a \emph{boundary condition} on the
horizon.

In the sequel, whenever we use the phrase ``Standard Model", it refers to any kind of quantized field theory suitable for describing relativistic
quantized particles, with gravity added perturbatively, if at all.

\newsec{\(U(1)\)-vector fields}\label{vector.sec} Consider a \(U(1)\) gauge field, \(A_\m(x)\), in our Standard Model, see also Ref.\cite{GtH96}.
This case suits us well to illustrate our general procedure. In 3+1 dimensions, the Lagrangian is \be S=\int\dd^4x \LL(x)\ ;\qquad \LL(x)=-\quart
F_{\m\n}F_{\m\n} \ . \eel{MaxwL} Given the charge distribution \(\r_\inn(\tl\s)\) of the ingoing particles, we find them to generate a vector field
\(A_\m(x)\), which becomes singular on the past horizon. By limiting ourselves to this singular contribution, we select out the effects of the horizon
itself that substitute as a boundary condition relating in- and outgoing objects. The particles that went in long ago, affect the particles that emerge in
the late future. The \(U(1)\) field generated by them takes the form of a distribution along the past horizon, as in the case of the generation of
\v{C}erenkov radiation. Before and behind the past horizon, the vector field is a pure gauge, but on the horizon these pure gauge parameters \(\L(\tl\s)\)
are making a finite jump. This gauge jump is found to be \be\L(\tl\s)=-\int\dd^2\tl\s'\,f(\tl\s-\tl\s')\r_\inn(\tl\s')\ ,
\qquad\tl\pa_\s^2\,f(\tl\s)=-\d^2(\tl\s)\ . \eel{gaugejump} On the past horizon itself, the vector field is not a pure gauge. Therefore, the gauge jump
\eqn{gaugejump} has a physically relevant effect on any charged outgoing particle: it performs a gauge rotation of the form \be e^{iQ_\out\L(\tl\s)}\ .
\eel{gaugetrfout} In total, the combined wave function of all outgoing particles undergoes a gauge rotation \be
e^{-i\int\dd^2\tl\s\int\dd^2\tl\s'\,f(\tl\s-\tl\s')\r_\inn(\tl\s)\r_\out(\tl\s)} \ ,\eel{gaugetrtot} with which we should multiply the amplitudes
generated by the algebra \eqn{pucomm}-\eqn{gravalgebra}.

We now introduce `functional plane waves' in the \(\r\) variables: \be\bra\{\l(\tl\s)\}|\{\r(\tl\s)\}\ket=\NN e^{i\int\dd^2\tl\s\, \r(\tl\s)\l(\tl\s)}\ ,
\eel{rlampl} both for the in-states and the out-states, where \(\NN\) is a normalization factor. \(\l(\tl\s)\) is a `functional momentum' operator: \be
\l_{\inn\atop\out}(\tl\s)=-i{\pa/\pa\,\r_{\inn\atop\out}}\ , \eel{lmom} and it obeys \be \big[\r_\inn(\tl\s),\l_\inn(\tl\s')\big]&=&
\big[\r_\out(\tl\s),\l_\out(\tl\s')\big]\iss i\d^2(\tl\s-\tl\s')\ , \crl{rlcomm} \r_{\inn\atop\out}(\tl\s)\dd^2\tl\s&=& i\pa/\pa\l_{\inn\atop\out}
(\tl\s)\ , \eel{rdl}

Comparing \eqn{rlampl} with Eq.\eqn{gaugetrtot}, which we write as the amplitude \(\bra\r_\out|\r_\inn\ket\), one gets \be
\l_\inn(\tl\s)&=&-\int\dd^2\tl\s'\,f(\tl\s-\tl\s')\r_\out(\tl\s')\ , \crl{rlinout} \l_\out(\tl\s)&=&\int\dd^2\tl\s'\,f(\tl\s-\tl\s')\r_\inn(\tl\s') \
,\eel{rloutin} or \be \r_\inn(\tl\s)=-\tl\pa_\s^2\,\l_\out\ , \qquad \r_\out(\tl\s)=\tl\pa_\s^2\,\l_\inn\ ;\crl{rinlout}
\big[\,\r_\inn(\tl\s),\,\r_\out(\tl\s')\,\big]=-i\tl\pa_\s^2\d(\tl\s-\tl\s') \ . \eel{inoutcomm} Note, that electric charge is quantized in multiples of
the electric charge unit \(e\). Therefore, the phase field \(\l(\tl\s)\) is only well-defined \emph{modulo} \(2\pi/e\). Thus, the physically relevant
field in Eq.~\eqn{rlampl} is \(e^{ie\l(\tl\s)}\,\), rather than \(\l(\tl\s)\).

This is the main part of, but not yet the complete algebra generated on a flat horizon by a \(U(1)\) gauge field whose Lagrangian in Minkowski space-time
is \eqn{MaxwL}. Note that we have no local gauge invariance on the horizon. This is because we assumed a pure vacuum state both in front of, and behind
the horizon. Those vacua are described by the preferred gauge choice for which \(A_\m(x)=0\). All we have is a global symmetry,
\(\L(\tl\s)\ra\L(\tl\s)+\L_0\), where the constant field \(\L_0\) is the generator. We return to this subject in Section \ref{transverse.sec}.

For future use, it is important to find the relation between the algebra \eqn{rlinout}---\eqn{rloutin} and the Lagrangian \eqn{MaxwL}. We write\fn{Sign
conventions in our publications are not uniform. Our present summation convention is \(X_\m Y_\m=X_+Y_-+X_-Y_++\tl X\tl Y\).} \be \LL(x)=-\quart
F_{\m\n}F_{\m\n}-J_\m(x)A_\m(x)\ , \eel{MaxwLJ} and, concentrating first on the contribution of the ingoing particles, \be J_+(x)=\r_\inn(\tl\s)\d(x^+)\
,\qquad J_-(x)=\tl J(x)=0\ . \eel{insource} In the Lorentz gauge, the vector field obeys \be\pa_+ A_-+\pa_-A_++{\tl\pa}\tl A&=&0\ \nn
2\pa_+\pa_-A_\m+{\tl\pa}^2 A_\m&=& J_\m\ , \eel{Maxwequ} so that \be A_+(x)&=&-\d(x^+)\L(\tl\s)\ , \qquad A_-=\tl A=0\ ,\nn \L(\tl\s)&=&
\int\dd^2\tl\s'\,f(\tl\s-\tl\s')\r_\inn(\tl\s')\iss\l_\out(\tl\s)\ . \eel{AMaxw}

Let us remind the reader how to go from here to the expression \eqn{gaugetrtot}. In the above solution, the vector field vanishes both before and behind
the horizon, so in both these regions, charged particles behave as in the vacuum. However, while crossing the horizon, the singular vector field
\eqn{AMaxw} affects their wave functions. To see the effect, a gauge transformation is performed as described at the beginning of this section. This gauge
transformation is chosen to be \be\L_1(x)=\L(\tl\s)\th(x^+)\ .\eel{thetagauge} The vector field \(A_\m(x)\) then transforms into \be A'_\m(x)=
A_\m(x)+\pa_\m\L_1(x) = \tl\pa_\m\L(\tl\s)\th(x^+)\ , \eel{regularvector} so that the singular expression \eqn{AMaxw} cancels out; only the transverse
derivatives of the gauge function survive. In this gauge, the charged fields obey regular field equations, so that they are continuous across the horizon,
and therefore, in this gauge, the outgoing wave functions \(\j_\out(x)\) pass through the horizon unaffected. But, to see the effects of the gauge field
\eqn{regularvector}, it is better to transform back: \be
\j_\out(x)\ra\j_\out(x)e^{-i\,q_\out\L(x)}=\j_\out(x)e^{-i\,q_\out\int\dd^2\tl\s'f(\tl\s-\tl\s')\r_\inn(\tl\s')}\ ,\eel{newwave} where \(q_\out\) is the
charge of an outgoing particle. If now the outgoing particles are assumed to generate a charge distribution \(\r_\out(\tl\s)\), then Eq.~\eqn{gaugetrtot}
is obtained.

\newsec{Scalar fields}\label{scalar.sec}

Consider now a set of scalar (or pseudo-scalar) fields \(\F^i(x)\) in the Standard model. The Lagrangian may be taken to be
\be\LL(x)=-\half(\pa\F^i)^2-V(\mathbf\F)\ . \eel{scalarL} Their effect on the algebra is nearly trivial, which is why we discussed the vector case first.
What has to be observed is, that a scalar field that lives on the plane \(x^+=x^-=0\) defining the horizon, remains invariant under a Rindler time boost,
\be x^+\ra \l\, x^+ ,\qquad x^-\ra x^-/\l\ . \eel{Rindlerboost} This implies that its value for the out-state is the same as for the in-state. The
``algebra" is therefore,
\be\F^i_\inn(\tl\s)&=&\F^i_\out(\tl\s)\ ; \\
\big[\F^i(\tl\s),\,\F^j(\tl\s')\big]&=&0\ . \eel{scalaralg}

This, however, is important. The scalar fields \(\F^i(\tl\s)\) give the horizon some conserved degrees of freedom; there is a local conservation law! This
local conservation law, however, differs from the local conservation laws in quantized gauge field theories; in a quantized gauge theory, the locally
conserved quantities always vanish. Here, they can take any set of values. The \(\F\) fields for an in-state simply take the same values as for its
corresponding out-state.

The role of the scalar field self-interaction, \(V(\mathbf\F)\), is the following. As long as these fields commute with all other operator fields on the
horizon, we must view them as Casimir operators. The quantity \be \mathbf{Z}(x^1,\,\cdots\,,\ x^n)=e^{\mathbf{W}(x^1,\,\cdots\,,\
x^n)}=\bra\mathbf\F(x^1)\,\cdots\,,\,\mathbf\F(x^n)\ket\ , \eel{scalarexpvalue} describes the correlations for the values of \(\mathbf\F\) as they occur
in the ``average" representation on the horizon. Here, \(\bra\mathbf\F(x^1)\,\cdots\,,\,\mathbf\F(x^n)\ket\) is the standard Green function, defined at
the points \(x^i\) on the horizon. Since these points are all spacelike separated, the quantity in question can be computed as if we were in Euclidean
space rather than Minkowski space-time. This Green function, of course, depends directly on the values of the scalar self-interaction \(V(\mathbf\F)\).

We imagine that the set of all states is finite (anticipating the fact that the entropy is finite), and that we must be describing an arbitrary, generic,
element of this set. The role played by the scalar field Lagrangian is to determine the \emph{weight} of the values for the scalar fields when averaging
over all black hole states. This is important, when we will be describing the BEH mechanism, see Section \ref{Higgs.sec}.

If the scalar field \(\mathbf\F\) transforms non-trivially under some symmetry transformation, be it a global one or a local one, then this symmetry may
appear to be explicitly broken at the horizon, since the field takes a fixed value. This, however, can be seen in a different light: since all values for
the field \(\mathbf\F\) occur in the complete set of all black holes, it may be more accurate to state that the black holes are degenerate under this
symmetry: the entire black hole transforms into another one under the symmetry transformation. If the symmetry group, or even its covering group, is not
compact, then a problem arises: the black hole appears to be in an infinite representation. Since its entropy is finite, such symmetries are not allowed.
This is Bekenstein's well-known argument\cite{Bekenstein} that black holes cannot observe the additive conservation laws associated to non-compact
symmetry groups.

When studying a finite region of the horizon, it is the surrounding scalar field that appears to break the symmetry explicitly. In this case, the scalar
field acts exactly as if it were a \emph{spurion}. Spurions\cite{spurion} were introduced in the '50s and '60s to describe explicit symmetry breaking such
as the breaking of isospin and flavor-\(SU(3)\).

\newsec{The BEH mechanism}\label{Higgs.sec}

Let us now introduce a complex Higgs field \(\f(x)\), replacing the Lagrangian \eqn{MaxwL} by \be\LL(x)=-\quart
F_{\m\n}F_{\m\n}-(D_\m\f)^*D_\m\f-V(\f,\f^*)\ , \eel{HiggsL} where \(D_\m=\pa_\m+ieA_\m\), and \(V\) must be gauge-invariant, typically
\(V=\half\l_H(\f^*\f-F^2)^2\), so that \(\bra\f\ket\ra F\).

 In Ref.~\cite{GtH96}, we learned how to include the effects due to the Higgs mechanism, henceforth referred to as the BEH mechanism\cite{BEH}. The
primary effect of the Higgs field is to add a mass term \(-\half M^2A_\m^2\) to the Lagrangian \eqn{MaxwL}, where \(M^2=2\l_HF^2\). This leads to a
similar mass term in \eqn{MaxwLJ}, so that the equation for the Green function \(f(\tl x-\tl x')\) in Eq.~\eqn{gaugejump} of our algebra gets modified
accordingly. This is because the Maxwell equations \eqn{Maxwequ} also receive this extra term. It was subsequently noted that, apparently, local gauge
invariance gets lost.

We can now understand in a more detailed way how this comes about. The scalar Higgs field, \(\f\), fluctuates around the dominating values \(|\f|=F\).
This breaks local gauge symmetry, but here, the breaking appears to be an explicit one, not a spontaneous one, since the field \(\L(x)\) is a scalar
field, not a vector field. Now, we connect this observation with our treatment of the scalar field in Section \ref{scalar.sec}. The scalar field acts as a
spurion. In Section~\ref{vector.sec}, Eq.~\eqn{Maxwequ} is replaced by \be 2\pa_+\pa_-A_\m+\tl\pa^2A_\m-2e^2\f^*\f A_\m=\pa_\m \pa_\n A_\n+J_\m+J_\m^\f\ ,
\eel{massMaxw} where \(J_\m^\f=-ie\f^*D_\m\f+ieD_\m\f^*\,\f\) is the scalar field's contribution to the current. We absorb this into the total current
\(J_\m\). As in the unbroken case in Section~\ref{vector.sec}, we now assume the total ingoing current to align along the past horizon, see
Eq.~\eqn{insource}. The \(x^-\)-dependence disappears, and Eq.~\eqn{gaugejump} is replaced by \be
(\tl\pa_\s^2-2e^2\f^*(\tl\s)\f(\tl\s))\,f(\tl\s,\,\s')=-\d^2(\tl\s-\s') \ . \eel{BEHjump}

The equations~\eqn{rlcomm}, \eqn{rlinout}, and \eqn{rloutin} are kept unchanged, but the function \(f(\tl\s,\,\tl\s')\) now obeys the new equation
\eqn{BEHjump}. It is noted that the entire scalar field, not just its vacuum expectation value, appears in the modified algebra.

The arguments at the end of Section~\ref{vector.sec} also remain unchanged: the question is, how does the singular vector field affect the outgoing wave
functions? Simply perform the gauge transformation \eqn{thetagauge}. The field equations are invariant \emph{provided we also gauge rotate the Higgs
field}; the scalar field is acting only weakly on the charged fields and so it cannot generate any singular behaviour. The gauge rotation, having the
theta jump \(\th(x^+)\), rotates the fields of the charged outgoing particles just as in the pure Maxwell case. It is important to realize that the gauge
rotation \eqn{newwave} is a transformation \emph{back} to the situation where all fields outside the horizon take their standard vacuum values, including
the scalar fields.

\newsec{The transverse gauge field components} \label{transverse.sec}
Having learned what the role is of scalar fields in our algebra, we now return to the transverse vector field components, \(\tl A(x)\). It is important to
note that, just as the scalar fields, these components of the vector field also remain invariant under a Rindler time boost. Clearly, they should be
treated in the same way as our scalar fields. For instance, the Higgs field must appear with the transverse covariant derivative in the Lagrangian
\eqn{scalarL}, and indeed, the transverse vector field plays a role when, in the previous Section, we rotate \(\f\) towards the positive real axis:
\(\f\ra F\) everywhere on the horizon. In a statistical sense, local gauge invariance is restored.

In summary: all field components that are invariant under the Rindler time translation, \be x^+&\ra& \l\, x^+\ , \nn x^-&\ra& x^-/\l\ , \eel{Rindlertime}
act as \emph{spurion fields} in the algebra; they are time translation invariant on the horizon. The fields \(A_+(x)\) are associated with the
\emph{in}-states, and \(A_-(x)\) with the \emph{out}-states, so that they generate the algebra \eqn{rinlout}, \eqn{inoutcomm}, but with modifications such
as \eqn{BEHjump}, due to the spurions.

\def\intt{\mathrm{int}}
\newsec{Quantum hair}
\def\inn{\mathrm{in}} \def\outt{\mathrm{out}} \def\SM{\mathrm{SM}}

At a first approximation, only the vector fields propagate properties of charged ingoing particles into those of charged outgoing ones. However, the
propagation of these vector fields depends on the scalars present. Since the scalar fields do not depend on Rindler time, we found that they are static,
and their action can be characterized as ``quantum hair". Not only the scalar fields are static, but also the transverse components of the vector fields.
Only the longitudinal gauge field components play the role of dynamically evolving objects: the \(A_+\) fields refer to the ingoing states, the \(A_-\)
fields to the outgoing ones (in Rindler light cone coordinates). Let us now summarize what we found in case of Abelian vector fields interacting with
scalars.

The resulting algebra can be read off directly from the Lagrangian. Let the scalar fields form a set of representations \(\mathbf\F\) of the local gauge
group(s) \(U(1)\). \be\LL^\SM(x)=-\quart F_{\m\n}F_{\m\n}-(D_\m\mathbf\F(x))^* D_\m\mathbf\F(x) - V(\mathbf\F,\,\mathbf\F^*)\ , \eel{LAbel} with the usual
definitions for an Abelian field strength \(F_{\m\n}\) and the covariant derivatives \(D_\m\) acting on the scalar fields. On the horizon, these scalar
fields take values \(\mathbf\F(\tl\s)\) that depend on the two transverse coordinates \(\tl\s=(\s^1,\,\s^2)\), but not on the Rindler time \(\t\).
Similarly, the \emph{transverse} components \(\tl A(\tl\s)\) do not depend on Rindler time. Being spacelike separated, these fields also all commute with
one another, so they simply play the role of distinguishing different `types' of black holes. Do note, that only (a part of) the Standard model
interactions were taken into account here; presumably, these `different' black holes will mix at a somewhat longer Rindler time scale, where gravitational
and other interactions cannot be ignored as it is done in this chapter. The black hole keeps the same haircut only during the time interval needed for a
distant test particle to approach the horizon closer than the cut-off scale of whatever Standard Model we have been considering.

The complete set of all possible black hole states form a distribution in the space of \(\mathbf\F\)- and \(\tl A\) values. The moments of this
distribution must be described by the quantum vacuum correlation functions \(\bra\mathbf\F(\tl\s^1)\cdots\,\tl A(\tl\s^n)\ket\).

The relation between the fields \(A_\pm(\tl\s)\) and the sources \(\r_\inn,\ \r_\outt\) are given by \emph{solving the classical Euler-Lagrange
equations}, using the given values for \(\mathbf\F(\tl\s)\,\) and \(\tl A(\tl\s)\), for the action \be
S=\int\dd^4x\LL^\SM(x)+\int\dd^2\tl\s\Big(\r_\inn(\tl\s) \l_\inn(\tl\s)-\r_\outt(\tl\s)\l_\outt(\tl\s)\Big)\ , \eel{SAbel} where \be
A_+(x)=\d(x^+)\l_\outt(\tl\s)\qquad \hbox{and}\qquad A_-(x)=\d(x^-)\l_\inn(\tl\s)\ . \eel{Epots}

Since \be\pa_-A_+=0\ ,\qquad\pa_+A_-=0\ , \eel{dpmA} the longitudinal derivatives of \(A_\pm\) will disappear from the kinetic part of \(\LL^\SM\). The
action for \(\l_\inn\) and \(\l_\outt\) will reduce to \be S&\ra&\int\dd^4x\Big(-\tl\pa A_+\tl\pa A_-
-A_+(q\mathbf\F)^*A_-q\mathbf\F-A_-(q\mathbf\F)^*A_+q\mathbf\F \Big)\nn && +\int\dd^2\tl\s\Big(\r_\inn\l_\inn-\r_\outt\l_\outt\Big) \nn &=&
\int\dd^2\tl\s\Big(-\tl\pa\l_\outt(\tl\s) \tl\pa\l_\inn(\tl\s)-2(q\mathbf\F)^* q\mathbf\F(\tl\s)\,\l_\outt(\tl\s)\l_\inn(\tl\s)\nn &&
+\r_\inn(\tl\s)\l_\inn(\tl\s)-\r_\outt(\tl\s)\l_\outt(\tl\s)\Big)\ , \eel{Lrhol} where \(q\) is the charge matrix for the scalar fields \(\mathbf\F\).
This gives the relations \be\r_\inn(\tl\s)=-\tl\pa_\s^2\l_\outt+\cdots,\qquad\r_\outt=\tl\pa_\s^2\l_\inn+\cdots\,, \eel{rdl1} where the ellipses refer to
the contributions of the spurion fields \(\mathbf\F\) on the horizon. The algebra, as derived in Section~\ref{vector.sec}, is now generated by the
commutation rules \be [ \r_\inn(\tl\s),\,\l_\inn(\tl\s') ]\iss [ \r_\outt(\tl\s),\,\l_\outt(\tl\s') ]\iss -i\d^2(\tl\s-\tl\s')\ . \eel{rlcomm1}

Naturally, we wish to generalize this result to the case of non-Abelian vector fields and fermions. However, blindly copying the above results to include
adjoint vector fields and scalars transforming in non-Abelian representations, would require the use of \emph{classical} Yang-Mills fields near a horizon,
while at the same time they carry charges. To be on the safe side, we concentrate first on the diagonal components of the vector fields only, writing all
charges as representations of the associated Cartan subalgebra of the local gauge algebra. In such a representation, the suppression of all off-diagonal
vector components of the gauge field gives us an Abelian vector theory with in addition magnetic monopoles.\cite{GtHAbelproj}

\newsec{Magnetic monopoles}  It turns out to be instructive first to consider this interesting intermediate case: adding magnetic charges to an
Abelian system. It is this theory that we now assume for the in- and outgoing particles at the horizon. In this chapter, the effects of extra scalar
fields will temporarily be ignored.

\def\El{\mathrm E} \def\Mg{\mathrm M} \def\ol{\overline} \def\Rre{\mathrm{Re}} We now have two kinds of sources, the electric
currents, of which we only take the lightcone components, \(\r_\inn^\El(\tl\s)\) and \(\r_\outt^\El(\tl\s)\), and the magnetic
currents, \(\r_\inn^\Mg(\tl\s)\) and \(\r_\outt^\Mg(\tl\s) \). Since magnetic monopoles generate vector potentials with Dirac
strings attached, we temporarily switch from vector potential notation to field strength notation. The Maxwell field due to the
ingoing electrically charged particles is read off from Eq.~\eqn{Epots}: \be
F_{i+}&=&\pa_iA_+\iss\d(x^+)\pa_i\l^\El_\outt(\tl\s)\ ,\qquad \tl\pa^2\l^\El_\outt\iss -\r^\El_\inn\ ,\nn F_{i-}&=&F_{ij}\iss 0\
, \eel{EFields} where, as in the rest of this paper, latin indices \(i,\,j,\dots\) refer to one of the two transverse indices
only. We write this as \be F_{i+}(\tl\s)=-\d(x^+)\tl\pa^{-2}\pa_i\r_\inn^\El(\tl\s)\ . \eel{EFieldsin} The Maxwell field due to
an ingoing magnetically charged particle is the EM-dual of that: \be F_{\m\n}^\Mg&\ra&\half\e_{\m\n\a\b}F_{\a\b}^\El\ ;\qquad
F_{i+}^\Mg\ \ra\ \e_{i+j-}F_{j+}\iss\e_{ij}F_{j+}\ ,\qquad\hbox{therefore}\nn
F_{i+}&=&-\d(x^+)\tl\pa^{-2}\Big(\pa_i\r^\El_\inn(\tl\s)+\e_{ij}\pa_j\r^\Mg_\inn(\tl\s)\Big) \ ,\quad F_{i-}\iss F_{ij}\iss 0\ .\
{\ }\eel{EMFields}

To find the effect of this field on the wave functions of outgoing charged particles, we have to write this Maxwell field in terms of a vector potential,
\be F_{i+}= \eth_iA_+\ . \eel{vecpot} The magnetic charges will generate Dirac strings in this vector potential, and this forces us to use a different
symbol \(\eth_i\) for partial differentiation here. As \(A_+\) is not unambiguous, the partial differential \(\eth_i\) may not obey the usual commutation
rules of ordinary derivatives. Let us introduce a convenient notation: \be z&=&\s^1+i\s^2\ ,\qquad\ol z\iss\s^1-i\s^2\ ;\nn A_+
&=&-(2\pi)^{-1}\d(x^+)\int\dd^2\tl\s\,\Rre\Big[\log(z-z')\bigg(\r_\inn^\El(z')+i\r_\inn^\Mg(z')\bigg)\Big] \ , \eel{EMvecpot} where the singularity of the
logarithm represents the Dirac string. To minimize the damage done by the Dirac string, we will later see that the source functions must be confined to be
of the type \be\r(\tl\s)=\sum_\a q_\a\d(\tl\s-\tl\s_\a)\ , \eel{deltas} where \(q_\a\) must be a multiple of some charge quantum \(e\) in the electric
case, and \(m\) in the magnetic case.

To find how this vector potential rotates the wave functions of the electrically charged outgoing particles, we have to write it
as the gradient of a gauge, \be A_+(x)=\pa_+\L(x)\ , \qquad \L(x)=\l_\outt^\El (\tl\s)\th(x^+)\ , \eel{gradgauge} and since the
wave functions rotate as \be \exp[i\int\dd^2\tl\s\,\l_\outt^\El(\tl\s)\r_\outt^\El(\tl\s)]\ , \eel{rotat} we again have
Eqs.~\eqn{rdl}, \eqn{rlcomm} and \eqn{rloutin}: \be \r_\outt^\El(\tl\s)\dd^2\tl\s&=&-i\pa/\pa\l_\outt^\El(\tl\s)\ , \nn
\big[\r_\outt^\El(\tl\s),\,\l_\outt^\El(\tl\s')\big]&=&-i\d^2(\tl\s-\tl\s')\ , \eel{rElEcomm} where \be \l_\outt^\El(z)\iss
-(2\pi)^{-1} \int \dd^2\tl\s\,\Rre\Big[\log(z-z')\bigg(\r_\inn^\El(z')+i\r_\inn^\Mg(z')\bigg)\Big]\ . \ {\ }\eel{lrEM}

We have similar equations replacing E by M, and furthermore, we assume \(\r_\inn^\El(\tl\s)\) to be independent of \(\r_\inn^\Mg(\tl\s)\), and similarly
\(\r_\outt^\El(\tl\s)\) to be independent of \(\r_\outt^\Mg(\tl\s)\), so that \be \big[\l_\inn^\El(\tl\s),\,\r_\inn^\Mg(\tl\s')\big]\iss
\big[\l_\outt^\El(\tl\s),\,\r_\outt^\Mg(\tl\s')\big]&=& 0\ , \nn \big[\l_\inn^\Mg(\tl\s),\,\r_\inn^\El(\tl\s')\big]\iss
\big[\l_\outt^\Mg(\tl\s),\,\r_\outt^\El(\tl\s')\big]&=& 0\ . \eel{mixedcomm} Combining these equations, one obtains
\be\big[\l^\El_\outt(\tl\s),\,\l^\El_\inn(\tl\s')\big]&=&(2\pi)^{-1}i\,\log|z-z'|\ , \nn
\big[\l^\El_\outt(\tl\s),\,\l^\Mg_\inn(\tl\s')\big]&=&-(2\pi)^{-1}i\,\arg(z-z')\ , \nn \big[\l^\Mg_\outt(\tl\s),\,\l^\El_\inn(\tl\s')\big]&=&
(2\pi)^{-1}i\,\arg(z-z')\ , \nn \big[\l^\Mg_\outt(\tl\s),\,\l^\Mg_\inn(\tl\s')\big]&=&(2\pi)^{-1}i\,\log|z-z'|\ . \eel{llcomm}

The relative signs here can be checked by noting the invariance \be \r^\El&\ra& \ \ \cos\th\,\r^\El+\sin\th\,\r^\Mg\ , \nn \r^\Mg&\ra& -\sin\th\,
\r^\El+\cos\th\,\r^\Mg\ , \eel{EMrotate} and similar rotations among the \(\l\) functions.

 In the mixed commutator expressions, the argument is defined \emph{modulo} \(2\pi\). Note however that, when two operators, \(A\) and \(B\) have a
c-number commutator, then \be e^{iA}\,e^{iB}=e^{iB}\,e^{iA}\ e^{-[A,\,B]}\ , \eel{CBH} Therefore, if we limit ourselves to the fields \(\exp(ie\l^\El)\)
and \(\exp(im\l^\Mg)\), such that \be e\cdot m=2\pi n\ ,\eel{Diracn} where \(n\) is an integer, then these fields commute unambiguously. This, of course,
is the Dirac condition on the magnetic charge quantum: it is the phase functions \eqn{rotat} that have to be defined unambiguously. We do need to define a
positive axis in the transverse space. This probably has to do with phase definitions for magnetic monopole wave functions.

We end this section with a number of formulae. Writing \be F_{i+}=\d(x^+)B_i^\inn(\tl\s)\ ,\eel{FB} one derives \be&&\pa_iB_i^\inn\iss -\r_\inn^\El\
,\qquad\e_{ij}\,\pa_i B_j^\inn\iss\r_\inn^\Mg\ ,\qquad\qquad\nn \hbox{so that}&\quad&\pa^2B_i^\inn\iss-\pa_i\r_\inn^\El-\e_{ij}\,\pa_j\r_\inn^\Mg\ .
\eel{divcurlB} We can now \emph{either} write \be B_i^\inn=\eth_i\l^\El_\outt\ ,\eel{BlEout} to describe the rotation of the electrically charged outgoing
particles, \emph{or} \be \,B_i^\inn=\e_{ij}\,\eth_j\l^\Mg_\outt\ , \eel{BlMout} to describe the outgoing magnetically charged particles. This gives the
commutation rules \be\big[B_i^\inn(\tl\s),\,\r_\outt^\El(\tl\s')\big]&=&i\pa_i\,\d(\tl\s-\tl\s')\ , \nn
\big[B_i^\inn(\tl\s),\,\r_\outt^\Mg(\tl\s')\big]&=&i\e_{ij}\,\pa_j\,\d(\tl\s-\tl\s')\ . \eel{Brhocomm} Furthermore, one has
\be\eth^2\l_\outt^\El=-\r_\inn^\El\ ,\qquad\eth^2\l^\Mg_\inn=-\r^\Mg_\inn\ , \eel{ddlr} but note that the operators \(\eth^2\) may not be inverted,
because the function \(\arg(\tl\s-\tl\s')\), being the angle formed by the 2-vector \(\tl\s-\tl\s'\) with respect to some fixed axis, obeys
\be\eth^2\,\arg(\tl\s-\tl\s')&=&0\ ,\crl{ddarg} \hbox{whereas}\qquad \pa^2\log|\tl\s-\tl\s'|&=&2\pi\,\d^2(\tl\s-\tl\s')\ .\qquad\qquad \eel{ddlog}

Expression \eqn{EMvecpot} can be cast into
\be\l_\outt^\El(\tl\s)&=&-(2\pi)^{-1}\int\dd^2\tl\s'\Big(\log|\tl\s-\tl\s'|\,\r_\inn^\El(\tl\s')-\arg(\tl\s-\tl\s')\r_\inn^\Mg(\tl\s')\Big)\
,\nn\l_\outt^\Mg(\tl\s)&=&-(2\pi)^{-1}\int\dd^2\tl\s'\Big(\log|\tl\s-\tl\s'|\,\r_\inn^\Mg(\tl\s')+\arg(\tl\s-\tl\s')\r_\inn^\El(\tl\s')\Big)\
,\nn\l^\El_\outt(\tl\s)+i\l^\Mg_\outt(\tl\s)&=&-(2\pi)^{-1}\int\dd^2\tl\s'\Big(\log|\tl\s
-\tl\s'|+i\arg(\tl\s-\tl\s')\Big)\Big(\r_\inn^\El(\tl\s')+i\r_\inn^\Mg(\tl\s')\Big) \nn && \eel{llrroutin}

We may use \be\eth_i\arg(\tl\s-\tl\s')=-\e_{ij}\pa_j\log|\tl\s-\tl\s'|\ ,\eel{paarg} and Eq.~\eqn{llrroutin}, or simply Eq.~\eqn{EMFields} to see that \be
B^\inn_i(\tl\s)=-(2\pi)^{-1}\int\dd^2\tl\s'\,\log|\tl\s-\tl\s'|\Big(\pa_i\r_\inn^\El(\tl\s')+\e_{ij}\pa_j\r^\Mg_\inn(\tl\s')\Big) \ . \eel{Binrho} With
Eqs.~\eqn{BlEout} and \eqn{BlMout}, \be[B_i^\inn,\,\r_\outt^\El]=i\pa_i\,\d^2(\tl\s-\tl\s')\ , \qquad
[B_i^\inn,\,\r_\outt^\Mg]=i\e_{ij}\,\pa_j\,\d^2(\tl\s-\tl\s')\ ,\eel{Brcomm} and taking Eq.~\eqn{Binrho} also to hold for the outgoing particles and
fields, one obtains
\be\big[B_k^\outt(\tl\s),\,B_i^\inn(\tl\s')\big]&=&-(2\pi)^{-1}i\int\dd^2\tl\s_1\log|\tl\s-\tl\s_1|\Big(\pa_i\pa_k+\e_{k\ell}\,\pa_\ell\,\e_{im}\pa_m\Big)
\d^2(\tl\s_1-\tl\s')\nn &=& -i\d_{ki}\,\d^2(\tl\s-\tl\s')\ . \eel{BBcomm}

This remarkably simple commutation rule should make it particularly easy to derive representations of our algebra, although we still have to take the
constraint \eqn{Diracn} into account. Remarkably, the simple commutator rule \eqn{BBcomm} would not hold had we left out the magnetic monopoles. This is
indeed a consequence of a peculiar feature of black holes: anything that can come out, will come out, so introducing the formal possibility of magnetic
monopoles implies a big modification of the operator algebra.

There is an important remark to be made. Eqs.~\eqn{BlEout} and \eqn{BlMout} imply that \(\l^\El_\outt\) and \(\l^\Mg_\outt\) are not independent. It may
seem mysterious how then to reconcile the non-vanishing commutator \eqn{rElEcomm} with Eqs.~\eqn{mixedcomm}. It seems that these equations must be
replaced by functions of \(\tl\s\) and \(\tl\s'\), which are extremely singular when \(\tl\s\) and \(\tl\s'\) coincide, but vanish as soon as
\(\tl\s\ne\tl\s'\). We must conclude that the points \(\tl\s\) where magnetic charges emerge or enter, never coincide with the points where electric
charges emerge or enter. Under that restriction, Eqs.~\eqn{mixedcomm} may be assumed to be correct.

\newsec{Towards Non-Abelian gauge fields. Conclusion.}\def\YM{\mathrm{YM}} One might expect the non-Abelian case to be described by charges \(\r^a\) and
gauge generators \(\l^a\). Naturally, we now could postulate the commutation rules as in Eq.~\eqn{rlcomm}: \be [ \r_\inn^a(\tl\s),\,\l_\inn^b(\tl\s')
]\iss [ \r_\outt^a(\tl\s),\,\l_\outt^b(\tl\s')]\iss -i\d_{ab} \d^2(\tl\s-\tl\s')\ . \eel{rlcommNA} However, we do note an imminent danger. We would like
to use these commutator equations to eliminate, for instance, \(\r_\inn^a(\tl\s)\) in the first of Eq.~\eqn{rlcommNA}. But, it had been assumed that
\(A_-^a\) was infinitesimal. This presumably means that we should take \(\l_\inn^a(\tl\s')\) infinitesimal, but this is not possible
--- it is an operator. Therefore, what really has to be done is a more general eikonal approximation involving non-Abelian gauge bosons. This is left
for future research.

What was found here is that we have sets of operators that live on the black hole horizon. Their commutation rules and eigenvalue spectra are controlled
by whatever our ``Standard Model" was taken to be. Scalar fields \(\mathbf\F\), as well as all other fields that are invariant under Rindler time
translations, such as the transverse vector field components, take values according to a distribution that must be the one generated by the 3+1
dimensional complete theory. They form what could be called `quantum hair'. The black hole haircut forms a (large) representation of the local and global
symmetries. If we single out one black hole in a given state, it may appear to violate the conservation laws dictated by this symmetry. However, as long
as the gauge group is compact, the conservation law is not violated: black holes simply have a pool of neighboring states to dump their excess charges in.

The other field components are non-commuting operators on the horizon. The complete set of black hole states must form a representation of this algebra.

The algebra of Section~\ref{intro.sec} only included the longitudinal gravitational forces, which certainly dominate at longer transverse distances. If,
however, we wish to include features where the length scales compare with the Planck scale, we definitely have to take the entire gravitational force into
account. Previous studies showed that this considerably modifies the algebra, but it was not clear exactly how to do this right, and the entropy area law
did not emerge naturally. At this point, it is still pure speculation that, if we manage to get the algebra correctly, its representations are expected to
display a discrete set of states, correctly reflecting the area law. Since the area law is generally believed to be hold, and since the algebra derived
here must be correct at large transverse distance scales, we believe that a proper extrapolation of this procedure should provide us with a consistent
picture. The attitude advocated in this paper is to learn how to set up these algebras step by step, so as to familiarize ourselves wit the rules and
procedures.

Superstring theories are generally expected to yield the entropy-area law more directly. Applying these to non-extreme horizons is difficult, however. One
might wonder whether the method advocated in this paper could be applied fruitfully to strings.

\end{document}